# Genomic and phenotypic characterisation of a wild Medaka population: Establishing an isogenic population genetic resource in fish

Mikhail Spivakov[1,6]*, Thomas O. Auer[2,7]*, Ravindra Peravali[3], Ian Dunham[1], Dirk Dolle[1,2], Asao Fujiyama[4], Atsushi Toyoda[4], Tomoyuki Aizu[4], Yohei Minakuchi[4], Felix Loosli[3§], Kiyoshi Naruse[5§], Ewan Birney[1§], Joachim Wittbrodt[2§]

[1]European Bioinformatics Institute (EMBL EBI), Wellcome Trust Genome Campus, Hinxton, Cambridgeshire, UK

[2]Centre for Organismal Studies (COS) Heidelberg, University of Heidelberg, Im Neuenheimer Feld 230, 69120 Heidelberg, Germany

[3]Karslruhe Institute of Technology, KIT, Karlsruhe, Germany

[4]Comparative Genomics Laboratory, Center for Information Biology, National Institute of Genetics, Yata 1111, Mishima, Shizuoka 411-8540, Japan

[5]National Institute for Basic Biology, NIBB, Laboratory of Bioresources, Okazaki, Japan

[6]Present address: Babraham Institute, Cambridge, UK

[7]Present address: Neuronal Circuit Development Group, Unité de Génétique et Biologie du Développement, U934 / UMR3215, Institut Curie, Paris, France

*These authors contributed equally to this work

§Corresponding authors

Email addresses:

    MS: Mikhail.Spivakov@babraham.ac.uk

    TOA: thomas.auer@curie.fr

RP: ravindra.peravali@kit.edu

ID: dunham@ebi.ac.uk

DD: ddolle@ebi.ac.uk

AF: afujiyam@gmail.com

AT: atoyoda@nig.ac.jp

TA : toaizu@nig.ac.jp

YM: yminakuc@nig.ac.jp

FL: felix.loosli@kit.edu

KN: naruse@nibb.ac.jp

EB: birney@ebi.ac.uk

JW: jochen.wittbrodt@cos.uni-heidelberg.de

# Abstract

**Background**

*Oryzias latipes* (Medaka) has been established as a vertebrate genetic model for over a century, and has recently been rediscovered outside its native Japan. The power of new sequencing methods now makes it possible to reinvigorate Medaka genetics, in particular by establishing a near-isogenic panel derived from a single wild population.

**Results**

Here we characterise the genomes of wild Medaka catches obtained from a single Southern Japanese population in Kiyosu as a precursor for the establishment of a near isogenic panel of wild lines. The population is free of significant detrimental population structure, and has advantageous linkage disequilibrium properties suitable for establishment of the proposed panel. Analysis of morphometric traits in five representative inbred strains suggests phenotypic mapping will be feasible in the panel. In addition high throughput genome sequencing of these Medaka strains confirms their evolutionary relationships on lines of geographic separation and provides further evidence that there has been little significant interbreeding between the Southern and Northern Medaka population since the Southern/Northern population split. The sequence data suggest that the Southern Japanese Medaka existed as a larger older population which went through a relatively recent bottleneck around 10,000 years ago. In addition we detect patterns of recent positive selection in the Southern population.

**Conclusions**

These data indicate that the genetic structure of the Kiyosu Medaka samples are suitable for the establishment of a vertebrate near isogenic panel and therefore inbreeding of 200 lines based on this population has commenced. Progress of this project can be tracked at http://www.ebi.ac.uk/birney-srv/medaka-ref-panel.

## Background

Defined genetic reference panels of inbred lines with divergent genotypes have often been exploited in genetics (reviewed in Flint and Mackay [1]). Broadly there are two approaches to creating such panels. The first involves crossing a small number of genetically distinct founders, followed by a number of interbreeding steps leading to an outcrossed population [2]. The outcrossed population is then inbred again to form recombinant inbred lines. The second approach involves capturing wild individuals from a population and inbreeding them to give near isogenic wild lines. Both methods result in a panel of inbred lines that has specific benefits for genetic mapping compared to using outbred individuals. First, as the same genotype can be produced multiple times from the panel, genotypic and phenotypic resources can be shared on the same panel between research groups and experiments. Second, the ability to make measurements from different individuals of the same genotype can overcome much of the non-genotypic study variance. Finally the environment can be systematically varied between many individuals of the same genotype, allowing dissection of interactions between genotype and environment, something which is not achievable in standard outbred populations. Recombinant inbred lines provide a more straightforward route to full characterization of the complete genetic components of a population, often relating phenotypes back to the founder strains. However, mapping resolution is restricted by the low number of recombinations available in the panel, and the overall diversity of the panel is limited by the diversity of the input founders, which can limit analysis of interesting traits. In contrast, near isogenic wild lines usually have both higher diversity of alleles (ideally with a consequent diversity in phenotypes) and have far better recombination patterns, allowing for the discovery of more genetic influences, and finer mapping of these loci. With the falling price of sequencing, determining the complete genotype for each line is no longer so onerous, leaving the inbreeding process itself and obtaining sufficient distinct lines to overcome multiple testing issues as the two major limitations of a near isogenic panel.

Recombinant inbred lines or near isogenic lines have been developed in many different species over the last 50 years. For example, in the maize genetics community the set of 302 IBM maize strains has been a cornerstone of both basic and applied research [3, 4]. In the *Arabidopsis* community the collection of 107 different wild accessions has allowed the exploration of the genetic determinants of a number of phenotypes and their relationship to the environment [5]. The development in *Drosophila* of both recombinant inbred lines [6] (>1700 lines) and near isogenic wild lines [7] allows genetic dissection of phenotypes coupled with the excellent transgenic and other resources in this organism. The yeast research community have used crosses between wild and laboratory strains [8], or surveys of wild species in related yeasts [9] to explore genotype to phenotype associations. In vertebrates the emphasis has been more on recombinant inbred lines. These include the BNxSHR cross in rats [10] and the Black6/DBA cross in mouse [11], both of which lead to a number of interesting traits being mapped in these species. The Mouse Collaborative Cross is the largest recombinant inbred line experiment undertaken in vertebrates [12], and is already showing promising results, although the mapping resolution will remain in the megabase range. So far the long generation times and difficulty in lab husbandry of wild indviduals has prevented, to our knowledge, the establishment of a near isogenic panel from the wild in any vertebrate species.

The model vertebrate Medaka (*Oryzias latipes*) is now being rediscovered beyond Japan for its developmental genetics, genomics and evolutionary biology [13, 14]. The physiology, embryology and genetics of Medaka have been extensively studied for the past 100 years. From 1913 onwards, Medaka was used to show Mendelian inheritance in vertebrates and in 1921 it was the first vertebrate in which crossing over between the X and Y chromosomes was detected [15, 16]. In Japan there are two divergent wild populations of Medaka separated by the Japanese Alps dividing the main island of Honshu (the Northern and Southern populations, Figure 1A) [17-20]. These two populations are not in sympatry (*i.e.*, do not overlap in the wild) and have many different phenotypic features; they can however produce fertile offspring when mated in the laboratory [21]. A critical feature of Medaka laboratory husbandry has been the routine inbreeding of wild individuals from the Southern Medaka population to isogenic strains pioneered by Hyodo-Taguchi in the 1980s [22, 23]. Some of these strains are now in their 80th brother-sister mating, and importantly, there are routine protocols for creating an inbred strain from the wild. At least 8 isogenic strains derived from single wild catches are available from the Medaka NBRP stock center [24]. Furthermore, the availability of standard transgenesis protocols [25], mutant lines [26], a 700MB reference genome sequence combined with a detailed linkage map [27], and tools for enhancer and chromatin analysis [28, 29] make Medaka a powerful vertebrate organism for developmental and molecular studies [30].

Here we characterize molecular, genetic and phenotypic variation in a single Southern Japanese population of Medaka to assess its suitability to found a near isogenic wild panel. For comparison we characterize a number of existing inbred strains, including the "reference" HdrR strain, and explore the population genomics of this species. In addition, we set the groundwork for routine high throughput phenotyping of Medaka strains to quantify morphometric features, which is one of many possible traits that can be measured in these vertebrates. On the basis of these results we have begun the inbreeding cycles of 200 lines for establishment of a Medaka inbred isogenic panel.

## Results

### Characterising Existing Isogenic Lines, including HdrR, by High Throughput Sequencing

The current Medaka reference assembly [27] is based on Sanger sequencing of the HdrR inbred strain derived from a Southern Japanese population. In order to verify and refine the reference genome sequence, we applied paired-end high throughput short read sequencing to genomic DNA from the HdrR inbred line, to a median depth of 144X (see Methods for details). Consistent with the high quality expected from the reference sequence, we identified 184,318 single nucleotides (0.026% of bases) unresolved in the reference assembly (labeled as 'N') and 102,432 single nucleotides (0.014 % of bases) that were discrepant between the reference genome assembly and the newly obtained sequence above our quality threshold. Further investigation showed that revision of discrepant bases in the reference sequence towards the newly determined sequence resulted in 89% of revised sites matching all individuals in an independently sequenced wild Southern Japanese population (described below), with 96% matching at least one of the analysed individuals in this population. In addition, revision of these discrepant positions resulted in a striking 2.5-fold increase in consistency with the presumed ancestral allele as determined by comparison to stickleback. Using direct sequencing after PCR amplification we were able to confirm the new sequence for 67.5% (27) of a set of 40 discrepant bases whereas the additional 32.5% (13/40) showed traces of both possible sequences and could not be disambiguated. These observations support the hypothesis that these divergent positions represent errors in the original reference sequence. The discrepancies span 490 genes and suggest 804 alternative amino acid sequences and 11 previously unmapped stop codons.

Our additional deep sequencing data provide an estimate of the original reference sequence error rate at around $2x10^{-5}$ assuming that there has been no sequence divergence in this strain since 2002. This is better than the estimated error rates for draft quality genome sequencing. Given the large depth of coverage of the newly obtained genome sequence, and the successful verification of a subset of bases in conflict, we accounted for revisions to the HdrR reference sequence based on the short read data in the subsequent analysis of all other samples in this study. Further refining the reference calls based on consistency with other inbred strains (sequenced as described below), we produced a revised reference assembly, which we submitted to the European Nucleotide Archive as accessions HF933207-HF933230 (see Materials and Methods for more detail).

We next used high throughput short read sequencing to characterise the genotypes of four additional inbred lines derived from three different locations: Northern Japan (HNI and Kaga, [22, 23, 31]), South Korea (HSOK, [24]) and China (Nilan, [24]). We discovered 15,343,105 single nucleotide variations within the four strains with divergence from the reference strain ranging from 1-1.18 % (Table 1), confirming the highly polymorphic nature of Medaka consistent with the broad geographical spread of its natural habitat [20]. Polymorphisms across the inbred lines were mapped to 16,938 out of 17,442 Medaka genes on chromosome contigs, of which ~34% lead to changes in amino acid sequence, including 175,820 missense mutations in 15,829 genes and 1682 nonsense mutations (Table 1). As expected, missense and particularly nonsense mutations were strongly depleted in favour of synonymous SNPs as compared to random expectation (34% vs 75.1% and 0.14% vs 4%, respectively, binomial test $p < 1 \times 10^{-300}$), presumably due to purifying selection. Of these nonsense mutations we confirmed 15 of 16 selected loci within the Kaga, HNI and Nilan strains by targeted PCR and sequencing. In addition we validated 9 out of 10 non-coding Kaga strain SNPs confirming the high reliability of our sequence data. The phylogenetic relationship between the strains based on the sequencing data (Figure 2A) was consistent with the geographical localization of the sampling sites of the inbred strains (Figure 1A) and earlier data based on a limited number of SNPs [27].

**Characterising Population Genetics Parameters of a Specific Southern Japanese Population of Medaka**

We set out to identify a genetically diverse, stable Medaka founder population for the establishment of a panel of further isogenic lines. For this purpose we sampled ponds in the Takashi Hongo and the Kiyosu areas in July 2010. We first analysed the mitotype of 50 and 109 individuals from the two sites respectively, as described by Takehana et al. [17]. The most likely source of population structure in these samples is human-mediated dispersal of fish, e.g. fish discarded from aquariums. We detected an unusual mitotype in the Takashi Hongo population, normally present only in Northern Japanese Medaka and which is a likely marker of human-mediated dispersal. Hence we decided to focus on the Kiyosu population in further analysis. We sequenced the highly variable D-loop region of the mitochondrial cytochrome B gene in 105 Kiyosu individuals and detected 8 different sequence variants. This degree of D-loop sequence diversity was comparable to an earlier study where Katsumura *et al.* [32] detected 11 sequence types in 124 individuals in the most diverse and 3 types in 159 individuals in the least diverse of three wild Medaka populations. Compared to the established Medaka inbred strains, the Kiyosu population clustered with the southern HdrR, H05 and HNCMH2 strains as expected (Supplementary Figure 1). We analysed microsatellite markers designed for 9 of the 24 Medaka linkage groups (Supplementary Table 1) and detected up to 21 different alleles within 105 individuals (Supplementary Table 2). An ideal founder population for establishing a Medaka population genomics panel would be without significant population structure. We calculated inbreeding coefficients (F) from these microsatellite allele frequencies. Negative values of F indicate excess heterozygosity (outbreeding) while positive values indicate heterozygote deficiency (inbreeding) compared with Hardy-Weinberg expectations. In the case of our founder population, F ranged from -0.188 to 0.203 with seven positive and two negative values and an average of 0.107. Based on this preliminary analysis we concluded that there is a tolerable degree of inbreeding within our population and decided to proceed with deeper analysis.

To analyse the Kiyosu population in greater detail, eight mating pairs and their first-generation progeny (i.e. 8 trios) were selected. The genomes of these individuals were profiled using high-throughput short read sequencing, with ~9X mean depth of coverage per line and compared to the Southern Japanese reference HdrR sequence. Approximately 23% of variants detected were monoallelic across Kiyosu, distinguishing this population from the reference HdrR strain. Using a trio-aware algorithm (TrioCaller [33]), we scored and phased the remaining 77% of variants, giving a total of 4,797,962 SNPs segregating within the Kiyosu population. Of these, 107,950 mapped to coding regions, with 60,333 synonymous, 47,617 non-synonymous and 611 nonsense substitutions. As in the inbred lines, the latter two mutation types were considerably depleted in favour of synonymous SNPs compared to random expectation (44.7% vs 75.1% and 0.06% vs 4%, respectively, binomial test $p<1 \times 10^{-300}$) as expected under purifying selection. Similar observations have been made recently in other population surveys of species such as Arabidopsis and human [34, 35].

This polymorphism set provides the opportunity for a far more powerful test of population structure than the microsatellite approach above. Clustering the sample genotype differences by distance metric we found that the parental genotypes were largely equidistant from each other in the tree without visible structure (Figure 2B). As expected, the wild individuals clustered closest to each other near the Southern reference strain HdrR collected in the same region of Japan confirming the results from directed genotyping above.

**Linkage Disequilibrium Estimation for Medaka**

Linkage disequllibrium (LD) is the deviation from independent segregation of alleles between loci, and is dependent on recombination rate and population history. The extent of LD is an important parameter for mapping genotype-phenotype associations within a population. We estimated the LD for the Medaka Kiyosu population from the 16 parental sample genotypes expressed as $r^2$ which is robust to smaller sample sizes and expresses the key relationship in association mapping. Figure 3A illustrates that median $r^2$ between pairs of loci gradually drops with increasing distance, reaching a minimum at a distance of around 12.5 kb. To compare this with human LD estimates obtained in the same setting, we selected 16 independent parents from trios sequenced by Complete Genomics [36]. In this human population $r^2$ stabilises at ~37 kb, which is ~3 times larger. Both Medaka and Human have substantially shorter LD than structured mouse populations. In the past, longer LD profiles were generally desirable in genetic panels for association analysis since this reduced the number of markers that needed to be typed to detect association. However, in the modern setting of inexpensive complete sequencing, a LD profile extending for shorter distances allows higher resolution and finer mapping of causal variants. For our Medaka data, ~85% of SNP pairs with $r^2 > 0.8$ mapped to the same gene, with ~37% mapping to the same exon, indicating that fine mapping in Medaka should largely be possible to the resolution of single genes and frequently even to a single exon. For comparison, only 56% and 29% of the equivalent SNP pairs mapped to the same gene and exon in the equivalent human population, respectively. We also estimated haplotype block size across 15 of the Medaka chromosomes using the method of Gabriel *et al.* [37]. The mean haplotype block size was 712 bp (median 259 bp) with a maximum of ~78 kb. Figure 3B shows the LD map of a representative region of chromosome 11. We conclude from the detailed genetic characterization of this Kiyosu population sample that these fish will be suitable for establishment of a population genomics resource.

**Phenotypic Analysis of Established Isogenic Strains**

The other important feature of a population panel is that it is expected to exhibit appreciable phenotypic variation across individual lines. To characterise one set of phenotypes we took advantage of the Southern inbred lines already in existence since these are likely to be representative of lines generated from the Kiyosu population, allowing us to easily characterise broad sense heritability of phenotypes. We used light microscope-based imaging combined with an automatic annotation algorithm to derive a number of morphometric features (Figure 4A) across four Southern and two Northern strains from both lateral and dorsal viewpoints. Reassuringly the body length measurement was almost perfectly correlated between the two viewpoints. As expected in a fish the majority of the variance between individuals correlates with body length, and so we normalised the remaining morphometric phenotypes by body length. Of the 7 phenotypes analysed, 4 show greater than 30% broad sense heritability i.e. the differences between strains explained 30% or more of the variance (Figure 4B). An example is shown in Figure 4C where a fish from the HdrR strain has substantially smaller eyes relative to body length than the Icab strain (compare box plot in Figure 4D). The broad sense heritability estimate is similar to analogous morphometric measurements between inbred mouse strains, suggesting that Medaka populations have similar levels of phenotypic variation as observed between laboratory mouse strains.

Many other established phenotypes that differ between Medaka strains have been observed during the prior century of research on this fish [18, 21, 23] and further work will be needed to examine the suitability of this population for each phenotype. Given the high diversity of alleles in the Southern population we can expect many phenotypes to have least some genetic variance in this population.

**Further Analysis of Japanese Medaka Populations**

The determination of polymorphism data in a wild Medaka population allows us to examine a number of questions in Medaka population genetics.

Phylogenetically, the Northern Japanese Medaka strains lie between the Southern Japanese and the Chinese and Korean Medaka populations which form an outgroup to the two Japanese clades [38]. There are several phenotypes unique to the Northern strains compared to all other Medaka, including a requirement for shallow water tanks for good viability, and differences in behavior, pigmentation (Figure 4B) and morphometric parameters (Supplementary Figure 2). Recently these divergent phenotypes have even led to suggestion that the Northern and Southern Medaka populations are distinct species [19].

**Probing for potential interbreeding between Medaka populations**

To clarify the genetic relationship between the Northern and Southern Japanese Medaka clades, we investigated potential wild interbreeding events. When interbreeding occurs, haplotypes will be exchanged between the populations, events that are referred to as introgression. In particular, older introgression events are expected to leave a recognizable footprint of longer, more similar haplotypes than expected between two divergent populations. Although these haplotypes might be challenging to identify individually, in a genome-wide comparison sufficient numbers of genome "histories" are sampled even in one individual to allow quite sensitive detection of introgression. A prominent example of the successful application of this approach is the careful observations of allele frequencies between Neanderthals and modern humans [39] and subsequently between Denisovans and moderns humans [40]. The same approach has recently been applied successfully to trace introgression in Sus (Pig) species in Indonesia [41]. These studies show that whole genome data can illuminate ancient introgression events which are not visible by methods typing more limited numbers of loci. In Medaka we employed the same approach by considering ancestral alleles identified by alignment to the stickleback genome sequence [42]. Under the null hypothesis of a complete population split the ancestral alleles are equally likely to be present in either of the two Northern strains. However, if one of the Northern strains had introgressed with the Southern strains, this symmetry would be split.

Using this approach, we found limited evidence of introgression between the Northern (HNI) and Southern Japanese samples. However, the observed effect was stronger in the Southern inbred HdrR strain than in the wild Southern population. Since these two samples share a high sequence similarity, we asked whether the weaker result for the wild population comes exclusively from alleles that it shares with HdrR, representing a "ghost population" artifact [43]. Indeed, when the analysis was limited to just SNPs that differentiate the wild population from HdrR, no evidence for introgression with HNI was obtained (see Supplementary Table 3). Similarly, we found evidence of introgression between the Chinese (Nilan) inbred strain and HdrR, but less so between Nilan and the wild Southern population (Supplementary Table 3). As these effects are predominantly observed for pairs of laboratory lines, they could reflect the population history of the wild popluations from which they were sampled, but may also be the results of historical laboratory breeding. Further, well sampled, wild individual sequences from Northern Japan, Korea and Eatern China would be needed to fully resolve this issue. At the same time, these results indicate that there was either extremely limited or no interbreeding between the Northern and Southern Medaka lineages at least to the point of the most recent common ancestor of the wild Kiyosu population or the Northern strains.

**The population history of Southern Medaka fish**

It has been shown in humans, pigs and giant pandas that it is possible to determine a population history in terms of effective population size from a single individual [44-46]. This approach exploits the presence of two haplotypes in each individual which each sample the population histories between the many different segments of the genome. A Hidden Markov model integrates the uncertainty of segments to provide a consistent view of the effective population size as a function of time. We performed a similar analysis on our Southern Medaka population, but for all the individuals in our wild catch sample. Figure 5 shows the distribution of population size estimates from the Kiyosu individuals. Except for the expected variability of estimates for very recent history, the population history profile was consistent between the individuals, and is a different shape from the other species analysed previously. This profile suggests a very large older population which went through a relatively recent bottleneck around 10,000 years ago (assuming a position-wise mutation rate of $2.5 \times 10^{-8}$ per generation, generation time = 1 year) followed by a modest expansion in recent history. In addition to its immediate significance for Medaka evolution, this result highlights the applicability and reproducibility of effective population size over time estimation for single-individual diploid genomes of model organisms determined using short-read sequencing.

**Evidence of recent selection in the Southern Medaka fish population**

Finally, we have used the wild genotypes to look for patterns of recent positive selection in the Southern population. An established footprint of recent selection is unusually long haplotypes of low diversity surrounding the allele under selection, while the alternative allele(s) are associated with haplotypes of a more typical length [47]. A metric termed Extended Haplotype Homozygosity (EHH) describes this effect for a given locus representing the probability that two randomly chosen chromosomes carrying the core haplotype of interest are identical by descent for the entire interval from the core region to any point $x$. EHH thus detects the transmission of an extended haplotype without recombination [47]. We applied an established statistical approach based on the ratio of EHH decay with distance (known as iHS for integrated Haplotype Statistic) for the ancestral and derived allele using the parental genotypes of the newly established Southern strain trios [48]. After applying stringent iHS filters (see Methods), we found that 192 SNPs, most of which map to 12 broad domains, showed evidence for recent positive selection of the derived haplotype (Table 2). As expected there were no cases with an unusually long ancestral haplotype at this cutoff. These domains contained 17 genes and in three cases the implicated SNPs mapped to their coding regions (Thioredoxin reductase 3 (defence against oxidative stress), Kelch-like 10, Solute carrier family 41), causing non-synonymous amino acid changes. Interestingly, nearly 20% of SNPs indicative of recent positive localised to highly conserved, potentially regulatory genomic regions (1.4-fold enrichment over random expectation, p=0.03, Supplementary Table 4). This indicates that recent selection events have resulted in potentially significant changes in the coding and non-coding sequence, paving the way for a further examination of their impact on the molecular function and physiology of the organism as whole.

# Discussion

Advances in genome sequencing have made it possible not only to expand the range of species for which reference genomes are available, but also to sample the genetic diversity of individuals within the same species. Here we have explored individual genetic diversity in the model vertebrate, Medaka (*Oryzias latipes*), using inbred strains derived from various geographically distinct populations and from wild catches obtained from a single Southern Japanese population.

Our primary motivation for this study was to characterize a single Medaka population from Southern Japan as a potential source for the first vertebrate near isogenic wild panel. The sampled individuals are genetically highly diverse as expected for a large population size. There is no discernible population structure in our sample which is beneficial for mapping by association. The relatively tight LD will prove invaluable for the isolation of causative variants with 85% of SNP pairs in LD (from this limited sample set) mapping to single genes, and 37% mapping to single exons. A larger panel will sample more recombination rates, and so we can expect this panel to nearly always have resolution to a single gene, and often to a single exon or regulatory region. As expected, the sample possess considerable phenotypic diversity, with the phenotypic differences similar in magnitude to differences between inbred Mouse strains.

During characterization of the progenitors of the panel, we also determined the sequence of a number of existing isogenic Medaka strains sampled from several wild populations. We are in the process of generating an improved reference assembly using the additional sequence generated in this project and other sources in the future. All the sequence data is freely available and the differences identified have been submitted to dbSNP and will be available on the Ensembl browser and other resources in the future. The list of variants provides a resource for current Medaka researchers, in particular those aiming to map and understand the different phenotypes present in these strains. As expected, amongst the wild catches and the inbred lines there are 2235 nonsense mutations representing natural knockouts of particular protein coding genes.

These initial data have allowed us to further explore Medaka population genetics. We lend genetic support to the current hypothesis that the Northern Medaka population is very likely to be a distinct species that has not significantly interbred with the Southern Medaka population since the Southern/Northern population split. The population history of Southern Medaka suggests a recent bottleneck from a substantially larger population around 10,000 years ago. Two events present themselves as possible explanations for this bottleneck. First, it is tempting to speculate that this might be associated with the development of rice cultivation on the Japanese islands. Second, there may have been transgression of Medaka populations after the last glacial age. 18,000 years ago, the sea level was about 120 m lower than at present and Japan was directly connected to the Asian continent. Around 10,000 years ago, the sea level rose and the present plain fields were beneath sea level. To investigate further we need to contrast the Southern Japanese Medaka population history to other histories from other geographic locations. The identification of three genes with putative amino acid changes due to positive selection is the start of exploring the impact of positive selection in Medaka. The Thioredoxin reductase-3, Kelch-like 10 and Solute carrier family 41 (Slc41A3) genes, involved in counteracting oxidative stress within the cell, spermatogenesis and magnesium transport respectively, might play fundamental roles for fitness and reproductive success. Intronic mutations were also found in other genes associated with membrane transport (Rims1, Slc39A10) and carbohydrate metabolism (Chst13), as well as signalling (Wnt4a) and chromatin (Histone 2AY), suggesting that they may well be under non-coding, potentially regulatory positive selection, consistent with the evidence from Stickleback [42].

## Conclusions

Overall this study shows that a near-isogenic wild Medaka panel has excellent potential for exploring vertebrate biology, analogous to the role of the Arabidopsis and Drosophila panels for plants and invertebrates, respectively. At the time of writing we have initiated the fourth generation of an inbreeding program for 200 independent Medaka lines derived from this Kiyosu population. On completion of this program these lines will be made broadly available to the community, with all data freely distributed. Of interest is that Medaka fish can be housed in similar facilities to Zebrafish, and similar phenotyping protocols often work. Progress of this project can be tracked at http://www.ebi.ac.uk/birney-srv/medaka-ref-panel/, and we welcome participants both from established Medaka and teleost laboratories and from the statistical genetics community.

## Methods

### Genomic DNA Extraction

Genomic DNA was extracted from one male specimen of each inbred strain at the age of about 8 weeks using Phenol/Chloroform/Isomylalcohol and RNA was removed using RNase A (Fermentas). For analysis of the wild catches DNA was extracted from caudal fins.

**Mitotyping**

Polymerase chain reaction - restriction fragment length polymorphism (PCR-RFLP) analysis of the mitochondrial cytochrome B gene was as described in Takehana *et al.*[17]. In brief a 1241-bp segment including the complete cytochrome B gene was amplified, amplicons were digested with five restriction endonucleases (*Hae*III, *Mbo*I, *Msp*I, *Rsa*I and *Taq*I) and mitotypes assigned by inspection.

**D-loop sequencing**

D-loop sequencing was as described in Katsumura *et al.* [32]. The 616-bp region of the mitochondrial D-loop was amplified from caudal fin clip genomic DNA by PCR using the following pair of primers (5' to 3'): Medaka_D-loop_F1: CCCAAAGCCAGGATTCTAA; Medaka_D-loop_R1: AACCCCCACGATTTTTGTC. Sequences were determined on both strands, and then trimmed to 508 bp for comparison.

**Design and analysis of microsatellite markers**

To identify potential microsatellite regions we compared the HdRr and the HNI genome using Sputnik [32] and identified regions with a maximal repeat unit length of 2 and the minimum length of SSR set to 20. We designed primers with primer3plus [49] to amplify a region of about 200 bp flanking the microsatellite repeat. We searched for 1 marker per chromosome but finally settled on 9 high quality microsatellite markers. Alleles were annotated from chromatograms after PCR amplification with fluorescently labelled oligonucleotides using GeneMarker software (Softgenetics). For a complete list of primers see Supplementary Table 1.

**Inbreeding coefficient calculation**

The Inbreeding coefficient F was calculated (in a single population) as $F = 1 - (H_{OBS} / H_{EXP})$ (equal to $(H_{EXP} - H_{OBS} / H_{EXP})$ where $H_{OBS}$ is the observed heterozygosity and $H_{EXP}$ is the expected heterozygosity calculated on the assumption of random mating.

**Short-read sequencing, reference assembly "patching" and SNP calling**

Eight paired-end libraries (insert size 300bp) and 3 mate-pair libraries (insert size 3K) from the HdrR reference strain were sequenced using the Illumina HiSeq machine to a median coverage of 144X. Reads were aligned using Bowtie2 [50] and SNPs called using SAMtools [51]. Raw sequence reads and annotations were submitted to DDBJ (accession DRA000588). The HdrR reference genome sequence was 'patched' using base differences passing a quality threshold of 100 and with additional bases not defined in the reference (*i.e.* Ns). However, when using this patched sequence as reference for calling SNPs in other inbred lines, we noticed that a small number of SNP calls (~0.14–0.2% total) were consistent with the base identified from the short-read HdrR sequence at quality scores below this threshold. Using an unthresholded HdrR dataset as the reference resolved this problem, albeit resulting in a marginal addition of 0.01%–0.017% SNPs that were consistent with the original reference. On the balance of the false-positive and false-negative rates estimated this way, we opted for using the unthresholded 'patched' reference for calling SNPs in all other strains. Furthermore, we supplemented the final patched sequence with 89,227 bases that were consistent with the other four inbred lineline sequences and submitted the final sequence as a Third Party Annotation record to the European Nucleotide Archive (HF933207-HF933230). For 41 loci, SNPs were validated by direct sequencing after PCR amplification using the primers listed in Supplementary Table 5. Samples from the other isogenic strains were sequenced using a combination of paired-end and mate-pair Illumina sequencing to a median coverage of 108-125X. Reads were aligned using Bowtie2 and SNPs called using samtools based on the patched assembly. Refer to Supplementary Table 6 for details.

24 wild individuals of the Southern population in 8 Mother-Father-Offspring Trios were sequenced using paired-end sequencing to a median coverage of 9X. The sequence was aligned to the patched reference sequence and SNPs were called using samtools. Alleles segregating within the population were then phased and annotated taking into account family structure using TrioCaller [33]. SNPs were filtered by the empirical $r^2$ between inferred and expected values ($r^2 >= 0.6$) and the observed Mendelian error (ERATE <= 0.1). Raw reads were submitted to ENA (accession ERP001016).

**Phylogenetic analysis**
Phylogenetic relationships between the wild Southern population, five inbred strains and stickleback were assessed using a neighbour-joining algorithm based on Kimura 2-parameter distances implemented in PHYLYP [52]. Genotypes for the wild population were called using a "majority vote" across the 48 haplotypes. Differences between Medaka samples and stickleback were too large to estimate distances. Hence the length of the stickleback branch shown in Figure 2 is an underestimate. The tree was rooted using the midpoint method and plotted using the T-REX online tool [53]. The relationships between the wild samples and the reference strain (Figure 2B) were analysed in R using a hierarchical clustering with Gower distances [54] based on the samtools genotypes.

**Estimation of linkage disequilibrium**

Linkage disequilibrium ($r^2$) was computed from the TrioCaller SNPs from the 16 Wild Southern trio founders as well as in 16 parents from human trios sequenced and annotated by Complete Genomics [36]. Computations were performed using VCFtools [55] with the following options: --ld-window-bp 50000, --max-alleles 2, --min-alleles 2, --min-r2 0.001 --geno 0.8. Haplotype blocks in the 16 Medaka founders were called using HaploView [56] using the default parameters for the method described by Gabriel *et al* [37] . Trials involving varying the minor allele frequency cut-off and the fraction of informative markers required to be in strong LD indicated that block size was relatively stable to these parameters (results not shown).

**Population size history analysis**

To estimate changes in population size over time, we used the method of Li and Durbin [44], implemented in the software package psmc. The following options were used with the psmc executable: -N25 -t15 -r5 -p "4+25*2+4+6", ensuring that the predicted number of recombinations occurring in each specified time interval were sufficiently high (>10) to prevent over-fitting. The output was scaled assuming a doubling time of 0.67 years per generation and a mutation rate of 2.5 x $10^{-8}$. Results for all individuals were combined by binning the results into $\log_{10}$ time intervals of length 0.1.

**Detection of signals of recent positive selection**

A method proposed by Voight *et al* [48] and implemented in the R package rehh [57] was used. The standardised output statistic (integrated haplotype score, iHS) was centered around zero, as expected, and was filtered using the following criteria to identify robust and consistent signals: (1) |iHS| >= 0.1 (corresponding to the top

0.05% of the distribution); (2) Within a 20kb window surrounding each SNP: (a) standard deviation (iHS) <= 0.34 across all SNPs (bottom 50% of the distirbution), (b) the number of SNPs with a iHS of at least 80% max(|iHS|) is at least 4, (c) the number of SNPs with a iHS of at least 80% max(|iHS|) consitutes at least 10% of all SNPs in the window. Thus clusters of SNPs with highly negative iHS values (indicative of derived alleles under recent positive selection) are detected. None of the positive iHS values (potentially corresponding to hitchhiker alleles or a "switched" directionality of selection, [48]) passed this filter.

**Phenotyping of morphometric features**

Morphometric features of four Southern (HdrR, Icab, HNCMH2, H05) and two Northern (Kaga, HNI) Medaka lines were analysed. All embryos were raised under identical environmental conditions. The embryos were imaged at 10dpf by mounting them in 85 % Glycerol. Imaging was performed using a Leica MZ 16 FA stereomicroscope with a Planapo 1.0x objective and 20x zoom factor. To extract morphological features, a manual and an automated algorithm were developed. While the manual algorithm allows the user to select features of interest, the automated algorithm automatically segments predefined landmarks. It is based on contour detection, morphological image filtering and connected component labeling, was implemented in Matlab and its accuracy was verified with the manual algorithm.

**Assessment of Heritability**

Broad sense heritability was calculated as the proportion of variance assigned to strain factors in an analysis of variance. The different morphometric measurements were divided by body length to remove the strongest source of variance; the ratios of the other measurements were analysed using the aov function in R, and the proportion of variance explained determined as the square of the residuals of the strain factor over the total sum of squares. Between 50 to 100 individuals were analysed in each strain.

**Data Submissions**

Whole genome sequencing of the 5 Medaka inbred strains was submitted to DDBJ under accession DRA000588. Whole genome sequencing of the eight Medaka wild catch trios from the Kiyosu population was submitted to the European Nucleotide Archive (ENA) under accession ERP001016. The revised genome sequence was submitted to ENA as accessions HF933207-HF933230. The SNPs discovered are currently being submitted to dbSNP and identifiers will be added here.

## Authors' contributions

MS carried out bioinformatics, phylogenetic and population genetic analyses, participated in sample collection and drafted the manuscript. TOA performed the molecular genetic studies and participated in sample collection and morphometric analysis. RP carried out the inbreeding and morphometric analysis. ID participated in the bioinformatic analysis and prepared the manuscript. DD participated in the SNP analysis and reference sequence patching and submitted the patched sequence. AF, AT, TA and YM sequenced the inbred strains. KN led sample collection and initial phenotyping. EB, JW, KN and FL conceived of the study, participated in its design and coordination and helped to draft the manuscript. All authors read and approved the final manuscript.

## Acknowledgements

We thank Stephen Fitzgerald and Javier Herrero for discussions on comparative alignements, Justin Paschall and Nima Pakseresht for asistance with database submissions, Hilary Browne at the Sanger Instute Sequencing Facility for sequencing of the inbred Trios, Tetsuaki Kimura for assistance with FISH sampling, Wei Chen for advice on TrioCaller, and the NBRP Core Facility Upgrading Program and Genome Information Upgrading Program for financial support of the resequencing of medaka inbred lines. RP and FL thank Nadeshda Wolf, Nadine Eschen, Natalja Kusminski and Nils Trost for assitance in inbreeding and morphometric studies.

# Figures

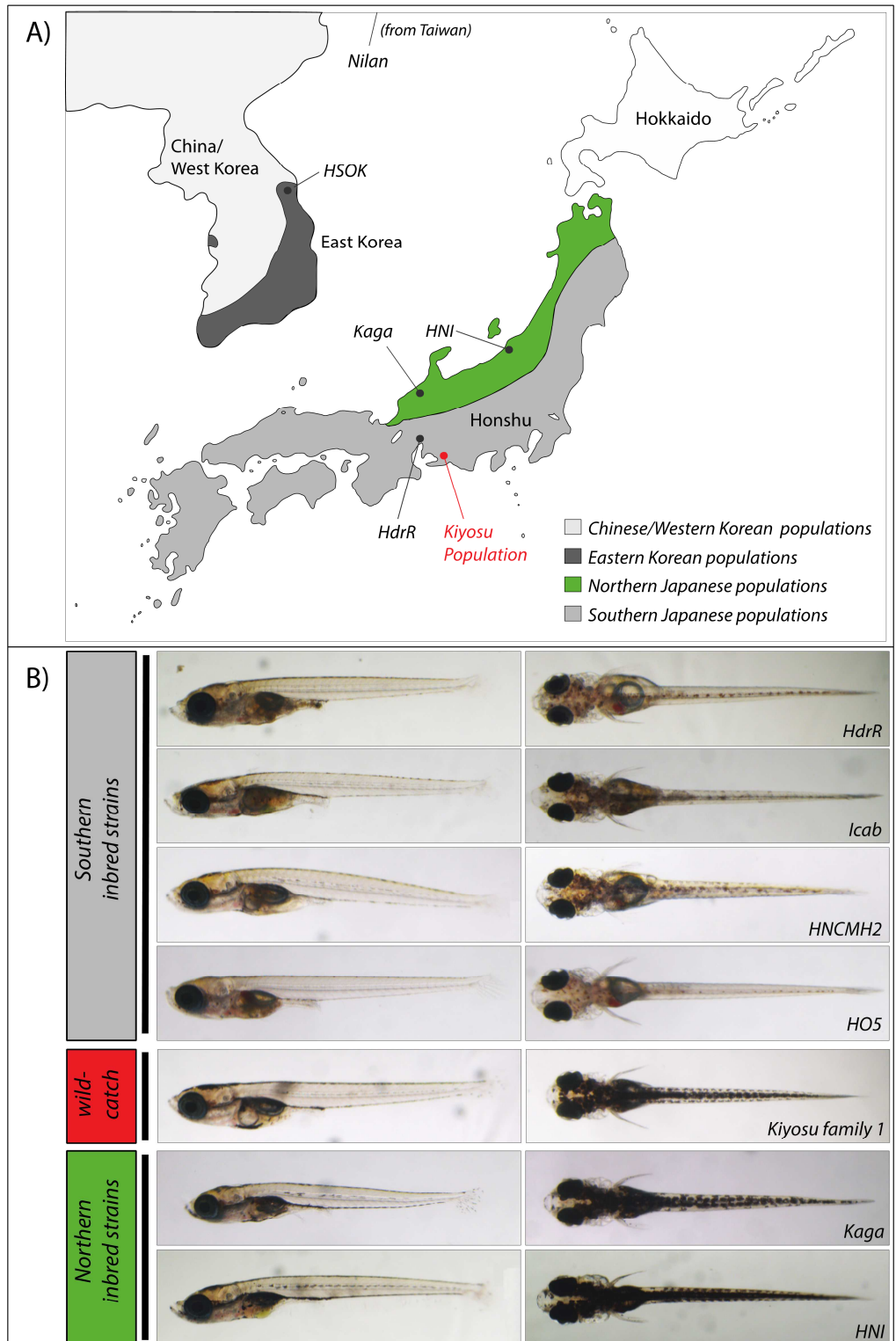

*Spivakov et al., Figure 1*

**Figure 1 - Geographical localization of the sampling sites for inbred stains and the Kiyosu population.**

A) Medaka consists of four major populations in East Asia. The Nilan strain comes from Ilan-City, Taiwan and is a representative of the Chinese/West Korean population; the HSOK strains from Sokcho-City, Gangweon-do, Korea and is a member of the East Korean population. The Japanese populations are divided by the Japanese Alps into Northern and Southern strains. While Kaga and HNI belong to the Northern populations, HdrR is the Southern reference strain that was used for the Medaka genome project. In red is the site in Kiyosu, where the founders for the inbreeding panel were sampled. B) Representative pictures of lateral and dorsal views of 10 days post fertilization old larvae of the different Japanese inbred strains. Icab, HNCMH2 and HO5 are further Southern Japanese inbred strains whose original sampling site is not shown in A).

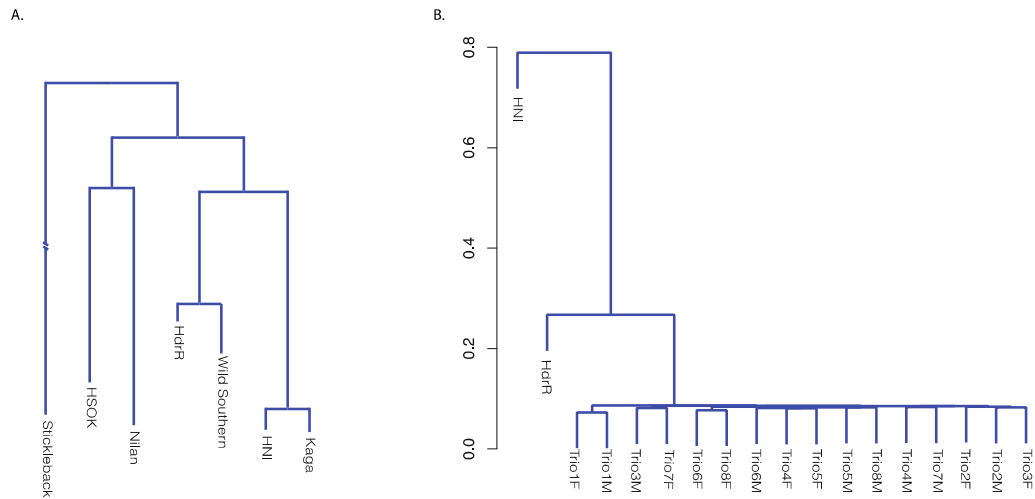

**Figure 2 – Phylogenetic Relationships of Medaka Strains and Kiyosu Population**

A) Phylogenetic relationships between the wild Southern population, five inbred strains and stickleback based on high throughput genome sequencing. PHYLIP clustering of inbred strains using a neighbour-joining algorithm based on Kimura 2-parameter distances is shown. Wild Southern is a synthetic sample based on the sequencing of the Kiyosu population and determining bases using majority vote across 48 haplotypes. Differences between Medaka samples and stickleback were too large to estimate distances. B) Hierarchical clustering with Gower distances of genotypes of Kiyosu population founder samples, and a representative Southern and Northern strains.

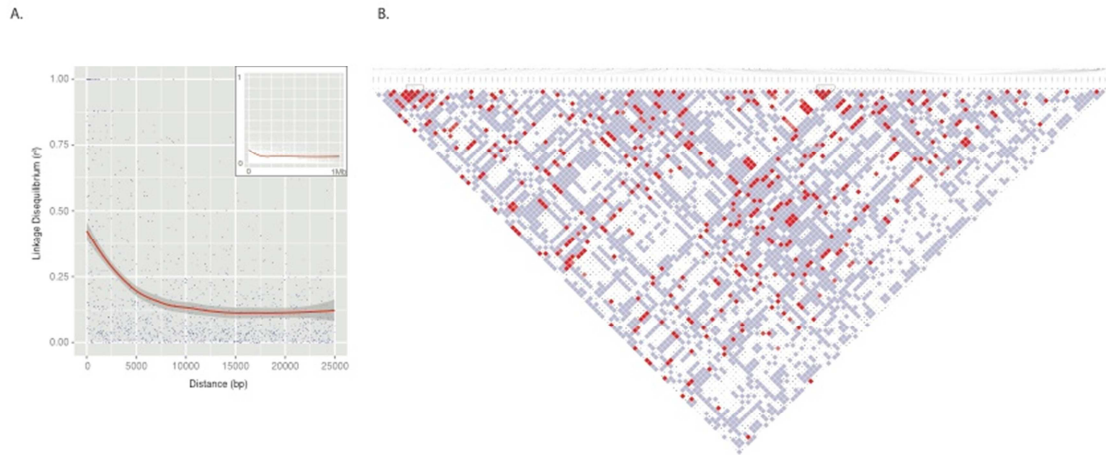

**Figure 3 - Linkage Disequilibrium in the Kiyosu Medaka Population.**

A) Decay of Linkage Disequilibrium (LD ) for a sample of Medaka SNPs. LD ($r^2$) was calculated for all SNP pairs in 1 Mb windows overlapping by 500 kb across Medaka chromosome 1, and $r^2$ was plotted against distance for a sample of 0.03 % (21976) of pairs. The main plot shows a blow up of distances between 0 and 25Kb, with data for pairs of SNPs up to 1 Mb apart shown in the inset region. Multiple sampling and analysis of other chromosomes gave similar results (not shown). B) Pairwise LD plot for a representative region of chromosome 11. Pairwise LD (D') for the 25 kb region 11:375000-399999 as displayed in Haploview is shown. The plot is colour coded as described for the standard LD colour scheme at http://www.broadinstitute.org/science/programs/medical-and-population-genetics/haploview/ld-display.

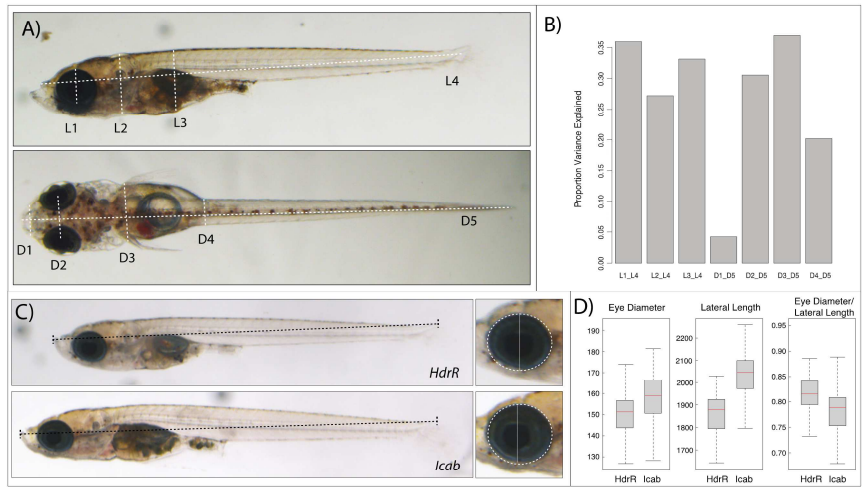

**Figure 4 - Morphometric analysis of Medaka inbred strains.**

A) Four lateral (L1-L4) and five dorsal (D1-D5) morphometric distances were extracted and analysed for each inbred strain. B) A bar chart showing the proportion of variance explained by the difference between strains as a fraction of the total variance. The variables are the measurements corrected by the appropriate body length measurement (L4 and D5 respectively). C) Example pictures showing substantial differences in lateral length and eye diameter between fish from the two Southern inbred strains HdrR and Icab. D) Box plots showing the distribution of eye diameter, lateral length and the ratio between eye diameter and lateral length for HdrR and Icab. Eye diameter and lateral length: y-axis = value in pixel. HdrR (n=70), Icab (n=78).

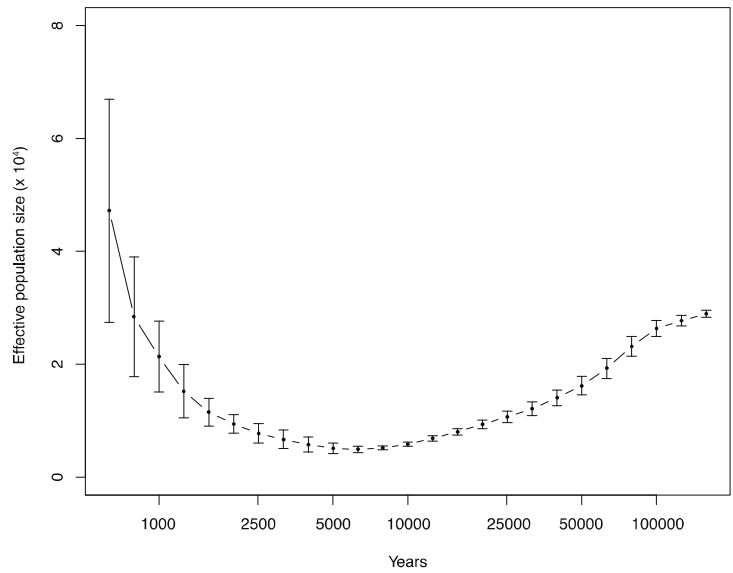

**Figure 5 - Distribution of population size estimates from the Kiyosu individuals.** Population size estimates over time were calculated using the psmc package for the founder individuals in the Kiyou wild catch as described in the Methods. The graph shows the distribution of estimates after combining by binning.

# Tables

**Table 1. Sequence variation in four Medaka inbred strains compared to the reference HdrR strain.**

Numbers are given for pairwise comparison with HdrR and compared to the 17,442 genes on chromosome scaffolds out of the 19,686 genes in the Medaka reference gene annotation.

| Strain | Variant Bases | Genes with Variants | Missense Bases | Nonsense Mutations |
|---|---|---|---|---|
| Nilan | 11,187,656 (1.60%) | 16,402 | 86,516 | 679 |
| Hsok | 11,832,789 (1.69%) | 16,603 | 83,241 | 755 |
| Kaga | 11,926,491 (1.7%) | 16,618 | 75,170 | 742 |
| Hni | 11,821,556 (1.69%) | 16,624 | 74,895 | 711 |
| **Total** | **26,371,226 (3.77%)** | **16,938** | **175,820** | **1682** |

**Table 2. Regions under positive selection in the Medaka genome.**
Chromosomal regions detected as under positivie selection are given along with details of the gene(s) contained within the regions and the locations of the indicative SNP(s).

| Chromosomal Region | Gene Symbol | Gene Description | Ensembl Gene ID | Location of indicative SNP(s) |
|---|---|---|---|---|
| 5:22601619-22654959 | SLC41A3 | Solute carrier family 41 | ENSORLG00000011740 | ORF (341: Met -> Val, 354 Leu -> Phe), 4 intronic |
| 5:22601619-22654959 | CHST13 | Carbohydrate (chondroitin 4) sulfotransferase 13 | ENSORLG00000011752 | 3 intronic, ORF (syn) |
| 5:23874411-24206161 | KLHL10 | Kelch-like 10 | ENSORLG00000011990 | ORF (100: Lys -> Asn; 353: Ala -> Thr), 1 intronic |
| 5:23874411-24206161 | TXNRD3 | Thioredoxin reductase 3 | ENSORLG00000011831 | ORF (457: Val -> Leu; syn), 4 intronic |
| 5:23874411-24206161 | BIN2 | Bridging integrator 2 | ENSORLG00000011869 | 4 intronic |
| 5:23874411-24206161 | WNT4 | Wingless-type MMTV integration site family member 4a | ENSORLG00000012025 | 2 intronic, 1 downstream |
| 4:5331132-5332602 | LRRC8C | Leucine rich repeat containing 8 family, member C | ENSORLG00000002275 | 3 intronic |
| 7:22770082-22774016 | RACGAP1 | Rac GTPase activating protein 1 | ENSORLG00000015540 | 1 upstream |
| 14:2110873-2114262 | H2AFY | H2A histone family, member Y | ENSORLG00000000850 | 7 intronic |
| 15:24520836-24521875 | COL19A1 | Collagen, type XIX, alpha 1 | ENSORLG00000010350 | 2 intronic |
| 15:7870807-8785145 | UBE2Z | Ubiquitin-conjugating enzyme E2Z | ENSORLG00000001771 | 5 intronic, ORF (syn) |
| 15:7870807-8785145 | RIMS1 | Regulating synaptic membrane exocytosis 1 | ENSORLG00000001748 | 5 intronic |
| 18:484757-484758 | Novel | Novel gene | ENSORLG00000002268 | 1 intronic |
| 21:2870759-2888920 | SLC39A10 | Solute carrier family 39 (zinc transporter), member 10 | ENSORLG00000008760 | 5 intronic |
| 1:9439527-9460390 | NA | NA | NA | gene desert |
| 9:14217964-14319610 | NA | NA | NA | gene desert |
| 11:25012896-25023149 | NA | NA | NA | gene desert |

# Additional files

**Supplementary Figure 1. Phylogenetic tree of Medaka inbred strains based on mitochondrial D-loop sequences.** The non-Japanese strains HSOK and Nilan form a separate cluster as does the Northern Japanese HNI strain. The sampled Kiyosu population is part of the Southern Japanese cluster as expected from the sampling site location.

**Supplementary Figure 2. Boxplots of measured morhpometric features in the different inbred strains.** A) Dorsal Features. B) Lateral Features. X-axis: Pixel values for each measurement. Grey shading: Southern inbred strains. Yellow shading: Nothern inbred strains.

**Supplementary Table 1. Microsatellite markers used.** Oligonucletoide primer sequences, genomic locations and amplicon lengths for the microsatellite markers analysed in the Kiyosu population sample.

**Supplementary Table 2. Microsatellite Alleles detected in Wild Kiyosu Population.** Microsatellite alleles were tested in 109 fish from the Wild Kiyosu Population. PCR amplification failed in 4 fish (samples 1, 5, 6, 7) giving allele results for 105 fish. For each microsatellite assay labelled as described in Supplementary Table 1, the two alleles identified in each Kiyosu fish assayed are listed.

**Supplementary Table 3. Introgression analysis.** Tests for introgression of the P3 strain into P1 or P2 strains that are more closely related with each other than to P1, using stickleback as the outgroup (O). D% is the introgression test statistic, on which the standard error (SE), 95% confidence intervals (CI) and Z-scores were estimated using block-wise jackknife [43].

**Supplementary Table 4. iHS analysis.** Numbers of SNPs showing evidence of recent positive selection in regions with different levels of evolutionary conservation versus those expected at random. Expected values, their standard deviations and p-values were estimated from permutation tests.

**Supplementary Table 5. Primers designed for SNP verification sequencing.** PCR amplification primers are given for the SNP resequencing amplicons. For the coding SNPs, the original reference sequence indicated a coding triplet, wheras a stop codon was indicated by our hight throughput sequencing. The status column indicates whether PCR amplicifcation and sequencing confirmed the new sequence.

**Supplementary Table 6. Details of sequence coverage for the additional inbred Medaka lines.** The table provides percent read alignment for each of the libraries sequenced for each inbred strain. Total coverage of the geonome and median read depth is given for the total sequence for all libraies from that strain.